\documentclass{aa}
\usepackage{graphicx}
\usepackage{txfonts}
\begin{document}
   \title{The frequency separations of stellar p-modes}


   \author{W. M. Yang,
          \inst{1, 2, 4}
          \and
          S. L. Bi
          \inst{3, 2}
          }

   \offprints{WuMing Yang}

   \institute{Department of Physics and Chemistry, Henan Polytechnic
              University, Jiaozuo 454003, China; \email{yangwuming@ynao.ac.cn}
             \and
             National Astronomical Observatories/Yunnan Observatory,
             Chinese Academy of Sciences, Kunming 650011, China
             \and
             Department of Astronomy, Beijing Normal University,
             Beijing 100875, China; bisl@bnu.edu.cn
             \and
             Graduate School of the Chinese Academy of Sciences,
             Beijing 100039, China\\
             }
  \date{Received January 11 / Accepted June 12}

   \abstract
    {}
    {The purpose of this work is to investigate the characteristics of a new
    frequency separation of stellar p-modes.}
    {Frequency separations are deduced from the asymptotic formula of stellar
    p-modes. Then, using the theoretical adiabatic frequencies of stellar model,
    we compute the frequency separations.}
    {A new separation $\sigma_{l-1 l+1}(n)$, which is similar to the scaled small
    separation $d_{l l+2}(n)/(2l+3)$, is obtained from the asymptotic formula of
    stellar p-modes. The separations $\sigma_{l-1 l+1}(n)$ and
    $d_{l l+2}(n)/(2l+3)$ have the same order. And like the small separation,
    $\sigma_{l-1 l+1}(n)$ is mainly sensitive to the conditions in the stellar
    core. However, with the decrease in the central hydrogen abundance of stars,
    the separations $\sigma_{02}$ and $\sigma_{13}$ deviate more and more from the
    scaled small separation. This characteristic could be used to extract the
    information on the central hydrogen abundance of stars.}
    {}

   \authorrunning{Yang and Bi}
   \titlerunning{Frequency separations}
   \maketitle

%

\section{Introduction}

   Helioseismology has proved to be a powerful tool for probing the
   structure of the Sun and has given us information on the interior
   of the Sun. The investigation of asteroseismology is stimulated
   by the success of the helioseismology and the verification of the
   solar-like oscillations in several stars, including $\alpha$ Cen
   A (Bouchy \& Carrier \cite{bou01}; Bedding et al. \cite{bed04}),
   $\alpha$ Cen B (Carrier \& Bourban
   \cite{car03}), $\eta$ Boo (Kjeldsen et al. \cite{kje95}), Procyon
   (Marti\'{c} et al. \cite{mar99}), and $\beta$ Hyi (Bedding et al.
   \cite{bed01}), etc.

   The goal of asteroseismology is to extract knowledge of the
   stellar internal structure that can be used to test and develop
   our understanding of stellar evolution from oscillation frequencies.
   The observation of solar-like oscillations is very difficult because
   of their small amplitude. Only a very limited
   number of modes ($\ell$= 0, 1, 2, 3) are likely to be observed in
   solar-like oscillations due to geometrical cancellation effects.
   How to extract the maximum information on the stellar
   internal structure from the limited modes is an important problem.

   The frequency separations including small separation and large separation
   have been successfully applied to extract
   information on stellar interior from oscillation frequencies in
   asteroseismology and have been investigated by many authors
   (Christensen-Dalsgaard \cite{chris84}, \cite{chris88}, \cite{chris93};
   Ulrich \cite{ulr86}, \cite{ulr88}; Gough \cite{gou87}, \cite{gou90b},
   \cite{gou03}; Gough \& Novotny  \cite{gou90a}; Roxburgh \& Vorontsov \cite{rox94a},
   \cite{rox94b}, \cite{rox00}, \cite{rox03};
   Audard \& Provost \cite{aud94}; Roxburgh \cite{rox05};
   Oti Floranes et al. \cite{flo05}). The usual frequency
   separations are the large separation defined by
   \begin{equation}
     \Delta _{l}(n)\equiv \nu_{n, l} - \nu_{n-1, l}
     \label{equla}
   \end{equation}
   and the small separation defined by
   \begin{equation}
     d _{l l+2}(n)\equiv \nu_{n, l} - \nu_{n-1, l+2}.
     \label{equsm}
   \end{equation}
   The second difference is given by (Gough \cite{gou90b}; Monteiro \& Thompson
   \cite{mon98}; Vauclair \& Th\'{e}ado \cite{vau04})
   \begin{equation}
     \delta _{l}(n)\equiv \nu_{n+1, l} + \nu_{n-1, l}-2\nu_{n,l},
     \label{equsec}
   \end{equation}
   and the difference is defined by Roxburgh (\cite{rox93},
   \cite{rox03}) as
   \begin{equation}
    d_{01}(n) \equiv (-\nu_{n, 1} + 2\nu_{n, 0}- \nu_{n-1, 1})/2.
   \end{equation}
   Moreover, the ratio of small separation to large separation,
   e.g.,
   \begin{equation}
     r_{l}(n)= \frac{d_{l l+2}(n)}{\Delta _{l}(n)},
     \label{ratio}
   \end{equation}
   was first pointed out to be independent of the outer layer of
   star and can be used to measure the stellar age by
   Ulrich (\cite{ulr86}).  Roxburgh \& Vorontsov (\cite{rox03}), Roxburgh
   (\cite{rox05}), and Oti Floranes et al. (\cite{flo05}) studied
   $d_{02}(n)/\Delta_{1}(n)$ and $d_{13}(n)/\Delta_{0}(n)$ in more
   detail and demonstrated that the ratio is essentially independent
   of the structure of the outer layer and is only determined by the
   interior structure.

   The frequency separations mentioned above have been used to
   diagnose the element diffusion in solar type stars (Vauclair \& Th\'{e}ado
   \cite{vau04}; Th\'{e}ado et al. \cite{the05}; Mazumdar \cite{maz05};
   Castro \& Vauclair \cite{cas06}) and the structure of stellar
   convective core (Roxburgh \& Vorontsov \cite{rox01}; Mazumdar et
   al. \cite{maz06}).

   Are there any other frequency separations? If there are other
   separations, what are their characteristics? In this paper,
   we focus mainly on investigating another frequency separation and
   some of its characteristics. In Sect. 2 we give the formulas for frequency
   separations. In Sect. 3 we present numerical calculation and results.
   Then, we discuss our results and conclude in Sect. 4.

\section{Asymptotic formula and frequency separations}
   The asymptotic formula for the frequency $\nu_{n, l}$ of a
   stellar p-mode of order $n$ and degree $l$ was given by Tassoul
   (\cite{tas80})
   \begin{equation}
     \nu_{n, l} \sim (n + \frac{l}{2} + \varepsilon) \nu_{0} -[Al(l + 1)
     - B] \nu_{0}^{2} \nu_{n, l}^{-1},
     \label{asym}
   \end{equation}
   for $n/(l + \frac{1}{2}) \rightarrow \infty$, where
   \begin{equation}
     \nu_{0} = (2 \int_{0}^{R} \frac{dr}{c})^{-1}
   \end{equation}
   and
   \begin{equation}
    A = \frac{1}{4\pi^{2} \nu_{0} } [\frac{c(R)}{R} -
    \int_{0}^{R}\frac{1}{r}\frac{dc}{dr}dr] ,
    \label{eqa}
   \end{equation}
   in which $c$ is the adiabatic sound speed at radius $r$, and $R$ is some
   fiducial radius of the star; $\varepsilon$ and $B$ are the quantities
   that are independent of the mode of oscillation but depend
   predominantly on the structure of the outer parts of the star;
   $\nu_{0}$ is related to the sound travel time across the stellar
   diameter; $A$ is a measure of the sound-speed gradient. In Eq. (\ref{eqa}),
   the integral is large compared with the surface term $c(R)/R$ (Gough
   \& Novotny \cite{gou90a}); therefore, $A$ is most sensitive to conditions
   in the stellar core and is invariant under homologous transformation.
   Consequently, the nonhomologous changes brought by the nuclear
   transmutation can be indicated by the variation in $A$
   (Gough \& Novotny \cite{gou90a}, \cite{gou03}; Christensen-Dalsgaard \cite{chris93}).
   Formula (\ref{asym}) is a second-order asymptotic description
   of low-degree p-modes under the Cowling approximation, in which the effects of
   gravitational perturbations are neglected. This formula is inaccurate except
   for very high frequencies (Roxburgh \& Vorontsov \cite{rox94a},
   \cite{rox94b}; Audard \& Provost \cite{aud94}).

   Using the definitions (\ref{equla}), (\ref{equsm}), and the asymptotic
   formula (\ref{asym}), Gough \& Novotny (\cite{gou90a}) obtained the large
   separation
   \begin{equation}
   \begin{array}{lll}
    \Delta_{l}(n)& = & \nu_{n,l}-\nu_{n-1,l}\\
    & = & \nu_{0}(\frac{\nu_{n-1, l} - [Al(l+1)-B]\nu_{0}^{2}\nu_{n,l}^{-1}}
    {\nu_{n-1, l}})^{-1} \\
    &\simeq &\nu_{0},
   \end{array}
   \label{larges}
   \end{equation}
   and the small separation
   \begin{equation}
   \begin{array}{lll}
    d_{l l+2}(n)& = & \nu_{n,l}-\nu_{n-1,l+2}\\
    &\simeq& \frac{[Al(l+1)-B]\nu_{0}^{2}(\nu_{n,l}-\nu_{n-1,l+2})}
    {\nu_{n,l}\nu_{n-1,l+2}}+\frac{2A(2l+3)\nu_{0}^{2}}{\nu_{n-1, l+2}}\\
    & \simeq & \frac{2A(2l+3)\nu_{0}^{2}}{\nu_{n-1, l+2}}\\
    &\simeq& \frac{2A (2l + 3)\nu_{0}}{n+l/2+\varepsilon}.
   \end{array}
   \label{smalls}
   \end{equation}
   The $\nu_{0}$ depends on the mean density of the star, hence on
   the mass and radius of the star. Thus $\Delta_{l}(n)$ puts a constraint
   on the radius of the star. The small separation $d_{l l+2}(n)$ is
   proportional to quantity $A$, therefore it is sensitive to the
   structure of the stellar core and the chemical compositions in the
   core. Thus it is related to the evolutionary stage (Gough \cite{gou87};
   Ulrich \cite{ulr86}).

   We define another difference in the frequencies,
   \begin{equation}
    \sigma_{l-1 l+1}(n) \equiv -\nu_{n, l-1}+2\nu_{n, l} - \nu_{n, l+1}.
   \end{equation}
   Using asymptotic formula (\ref{asym}), we can get
   \begin{equation}
   \begin{array}{lll}
    \sigma_{l-1 l+1}(n) &=& 2\nu_{n, l} - \nu_{n, l+1} - \nu_{n,l-1}\\
    &\simeq&-\frac{\nu_{0}}{2} +
    \frac{[Al(l+1)-B]\nu_{0}^{2}(\nu_{n,l}-\nu_{n,l+1})}{\nu_{n,l}\nu_{n,l+1}}
    + \frac{2A(l+1)\nu_{0}^{2}}{\nu_{n,l+1}}\\
    & & +\frac{\nu_{0}}{2} +
    \frac{[Al(l+1)-B]\nu_{0}^{2}(\nu_{n,l}-\nu_{n,l-1})}{\nu_{n,l}\nu_{n,l-1}}
    - \frac{2Al\nu_{0}^{2}}{\nu_{n,l-1}}\\
    &=&\frac{[Al(l+1)-B]\nu_{0}^{2}(\nu_{n,l}-\nu_{n,l+1})}{\nu_{n,l}\nu_{n,l+1}}
    -\frac{[Al(l+1)-B]\nu_{0}^{2}(\nu_{n,l-1}-\nu_{n,l})}{\nu_{n,l}\nu_{n,l-1}}\\
    & &+ \frac{2A(l+1)\nu_{0}^{2}}{\nu_{n,l+1}}
    - \frac{2Al\nu_{0}^{2}}{\nu_{n,l-1}}.\\
   \end{array}
   \label{halfsigs}
   \end{equation}
   Equation (\ref{halfsigs}) can be rewritten as
   \begin{equation}
   \begin{array}{ll}
   (\nu_{n,l}-\nu_{n,l+1})(1-\frac{[Al(l+1)-B]\nu_{0}^{2}}{\nu_{n,l}\nu_{n,l+1}})
   & \\
   +(\nu_{n,l}-\nu_{n,l-1})(1-\frac{[Al(l+1)-B]\nu_{0}^{2}}{\nu_{n,l}\nu_{n,l-1}})&=
    \frac{2A(l+1)\nu_{0}^{2}}{\nu_{n,l+1}}
    - \frac{2Al\nu_{0}^{2}}{\nu_{n,l-1}}.
   \end{array}
   \end{equation}
   Because $[Al(l+1)-B]\nu_{0}^{2}/(\nu_{n,l}\nu_{n,l\pm1}) \ll 1$,
   then
   \begin{equation}
   \begin{array}{lll}
    \sigma_{l-1 l+1}(n)& \simeq &\frac{2A(l+1)\nu_{0}^{2}}{\nu_{n,l+1}}
    - \frac{2Al\nu_{0}^{2}}{\nu_{n,l-1}}\\
    &= &\frac{2A\nu_{0}^{2}}{\nu_{n,l+1}}(1+\frac{l
    (\nu_{n, l-1}-\nu_{n, l+1})}{\nu_{n, l-1}}).
   \end{array}
   \end{equation}
   For the low-degree p-modes,
   $| l(\nu_{n, l-1}-\nu_{n, l+1})/\nu_{n, l-1}| < 1$, we can therefore
   get
   \begin{equation}
   \sigma_{l-1 l+1}(n)\approx\frac{2A\nu_{0}^{2}}{\nu_{n, l+1}}. \label{sigs}
   \end{equation}

\begin{figure*}
     \includegraphics[angle=-90, width=8cm]{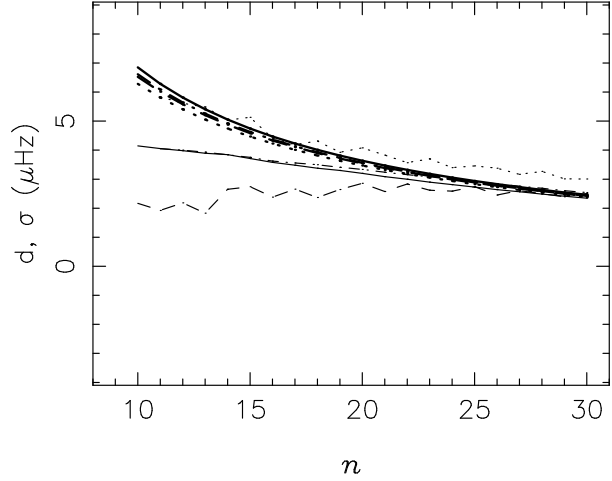}
     \centering
       \caption{Frequency separations for model M1.0 at the age of 4.5 Gyr.
       The solid line refers to $d_{02}(n)/3$. The dash-dot line shows
       $d_{13}(n)/5$. The dashed line indicates $\sigma_{02}(n)$.
       The dotted line corresponds to $\sigma_{13}(n)$. The data of bold
       lines are computed using the asymptotic formulas (\ref{smalls}) and
       (\ref{sigs}), but the data of other lines are computed using jig6 code
       (Guenther et al. \cite{gue92}).}
       \label{fig1}
   \end{figure*}

   On one hand, comparing Eq. (\ref{sigs}) with Eq. (\ref{smalls}), the
   difference $\sigma_{l-1 l+1}(n)$ and the scaled small separation
   $d_{l l+2}(n)/(2l+3)$ should have similar characteristics, which
   are proportional to quantity $A$ and sensitive to the conditions
   in the stellar core. The nonhomologous changes brought on by nuclear
   transmutation should thus be indicated by the variations in the scaled small
   separation and the difference $\sigma_{l-1 l+1}(n)$. On the other hand,
   Roxburgh \& Vorontsov (\cite{rox94a}, \cite{rox94b}) and Audard \&
   Provost (\cite{aud94}) have shown that the asymptotic formula (\ref{asym})
   is inaccurate except for very high frequencies. In Fig. \ref{fig1}, we
   compare the separations $d_{l l+2}(n)/(2l+3)$ and $\sigma_{l-1 l+1}(n)$
   obtained from the asymptotic formulas (\ref{smalls}) and (\ref{sigs}) with
   the separations obtained from the numerically computed frequencies.
   The discrepancy between the asymptotic and the numerical values is large,
   as has been shown by Roxburgh \& Vorontsov
   (\cite{rox94a}) and Audard \& Provost (\cite{aud94}).
   Equations (\ref{smalls}) and (\ref{sigs}) obtained from the asymptotic
   formula (\ref{asym}) are inaccurate. There may thus be misleading
   in the conclusion obtained from comparing Eq. (\ref{sigs}) with
   Eq. (\ref{smalls}). Moreover, the term
   $ l(\nu_{n, l-1}-\nu_{n, l+1})/\nu_{n, l-1}$ neglected
   in Eq. (\ref{sigs}) depends on degree $l$ and order $n$.
   Consequently, the difference $\sigma_{l-1 l+1}(n)$ should
   be somewhat different from the scaled small separation
   $d_{l l+2}(n)/(2l+3)$ and more dependent on degree $l$ than
   $d_{l l+2}(n)/(2l+3)$.

   The more accurate eigenfrequency equation was given
   by Roxburgh \& Vorontsov (\cite{rox00}, \cite{rox03})
   \begin{equation}
    2\pi T\nu_{n, l}=(n+l/2)\pi+\alpha(2\pi\nu_{n, l})-\varphi_{l}(2\pi\nu_{n,
    l}),\label{eigf}
   \end{equation}
   where $T=\int_{0}^{R}\frac{dr}{c}$ is acoustic radius, $\alpha(2\pi\nu)$
   the surface phase shift, and $\varphi_{l}(2\pi\nu)$ the internal phase shift
   that only depends on the interior structure of the star.
   Using Eq. (\ref{eigf}), Roxburgh \& Vorontsov (\cite{rox03}) get
   \begin{equation}
   \Delta_{l}(n)=\frac{1}{2T},\label{eigdelta}
   \end{equation}
   and
   \begin{equation}
   d_{l l+2}(n)=\frac{\varphi_{l+2}-\varphi_{l}}{2\pi T}.
   \label{eigd}
   \end{equation}
   Using Eq. (\ref{eigf}), we can get
   \begin{equation}
   \sigma_{l-1 l+1}(n)=\frac{\varphi_{l+1}+\varphi_{l-1}-2\varphi_{l}}{2\pi
   T}.\label{eigsigma}
   \end{equation}
   Comparing Eq. (\ref{eigd}) with Eq. (\ref{eigsigma}), one can find that
   the difference $\sigma_{l-1 l+1}(n)$ should be different from the small
   separation $d_{l l+2}(n)$ or the scaled small separation $d_{l l+2}(n)/(2l+3)$
   because they rely on the different internal phase shifts $\varphi_{l}$, which
   strongly depends on the degree $l$ (Roxburgh \& Vorontsov \cite{rox00}).
   However, if one assumes $\varphi_{l}\sim l(l+1)D_{\varphi}$, where
   $D_{\varphi}$ is a quantity determined only by the refractive
   properties of the stellar core (Roxburgh \& Vorontsov
   \cite{rox00}), one can get
   \begin{equation}
   d_{l l+2}(n)\sim\frac{(2l+3)D_{\varphi}}{\pi T},
   \label{eigd2}
   \end{equation}
   and
   \begin{equation}
   \sigma_{l-1 l+1}(n)\sim\frac{D_{\varphi}}{\pi
   T}.\label{eigsigma2}
   \end{equation}
   From Eqs. (\ref{eigd2}) and (\ref{eigsigma2}), we can find that
   the separations both $\sigma_{l-1 l+1}(n)$ and $d_{l l+2}(n)/(2l+3)$
   mainly depend on $D_{\varphi}$, and they should have some common
   characteristics. However, this conclusion can be obtained only under
   the approximation of $\varphi_{l}\sim l(l+1)D_{\varphi}$, which is
   inaccurate (Roxburgh \& Vonrontsov \cite{rox00}).

\section{Numerical calculation and results}

   We use the Yale Rotation Evolution Code (YREC7) to construct the
   stellar models in its nonrotating configuration. All models are evolved from
   the pre-main sequence to somewhere near the end of the main sequence. The
   newest OPAL EOS-2005\footnote{www-pat.llnl.gov/Research/OPAL/}
   (Rogers \& Nayfonov \cite{rog02}), OPAL opacity (Iglesias \& Rogers \cite{igl96}),
   and the Alexander \& Ferguson (\cite{ale94}) opacity for low temperature are used.
   Element diffusion is incorporated for helium and metals (Thoul et al.
   \cite{tho94}). The parameters of the models calculated are listed in Table
   \ref{tab1}. The mixing-length parameter $\alpha$ and hydrogen abundance are
   scaled to obtain the solar radius and luminosity, respectively, at the age of
   4.5 Gyr for model M1.0. Adiabatic oscillation frequencies of all models are
   computed using Guenther \& Demarque pulsation code jig6
   (Guenther et al. \cite{gue92}).

   \begin{table}
   \caption[]{Model parameters}
   \label{tab1}
   \centering
   \begin{tabular}{c c c c}
   \hline\hline
     parameter & M1.0 &  M1.1 & M1.2 \\
   \hline
       Mass     &1.0 $M_{\odot}$ &1.1 $M_{\odot}$  & 1.2 $M_{\odot}$ \\
     $\alpha$   & 1.720          & 1.720           & 1.720     \\
     $X_{0}$    &0.706           & 0.706           & 0.706    \\
     $Z_{0}$    &0.020           & 0.020           & 0.020    \\
   \hline
   \end{tabular}
   \begin{list}{}{}
     \item Note.--The $\alpha$ is the mixing-length parameter;
       $X_{0}$ and $Z_{0}$ are the initial hydrogen and metal
       abundance, respectively.
   \end{list}
   \end{table}

\begin{figure*}
     \includegraphics[angle=-90, width=9cm]{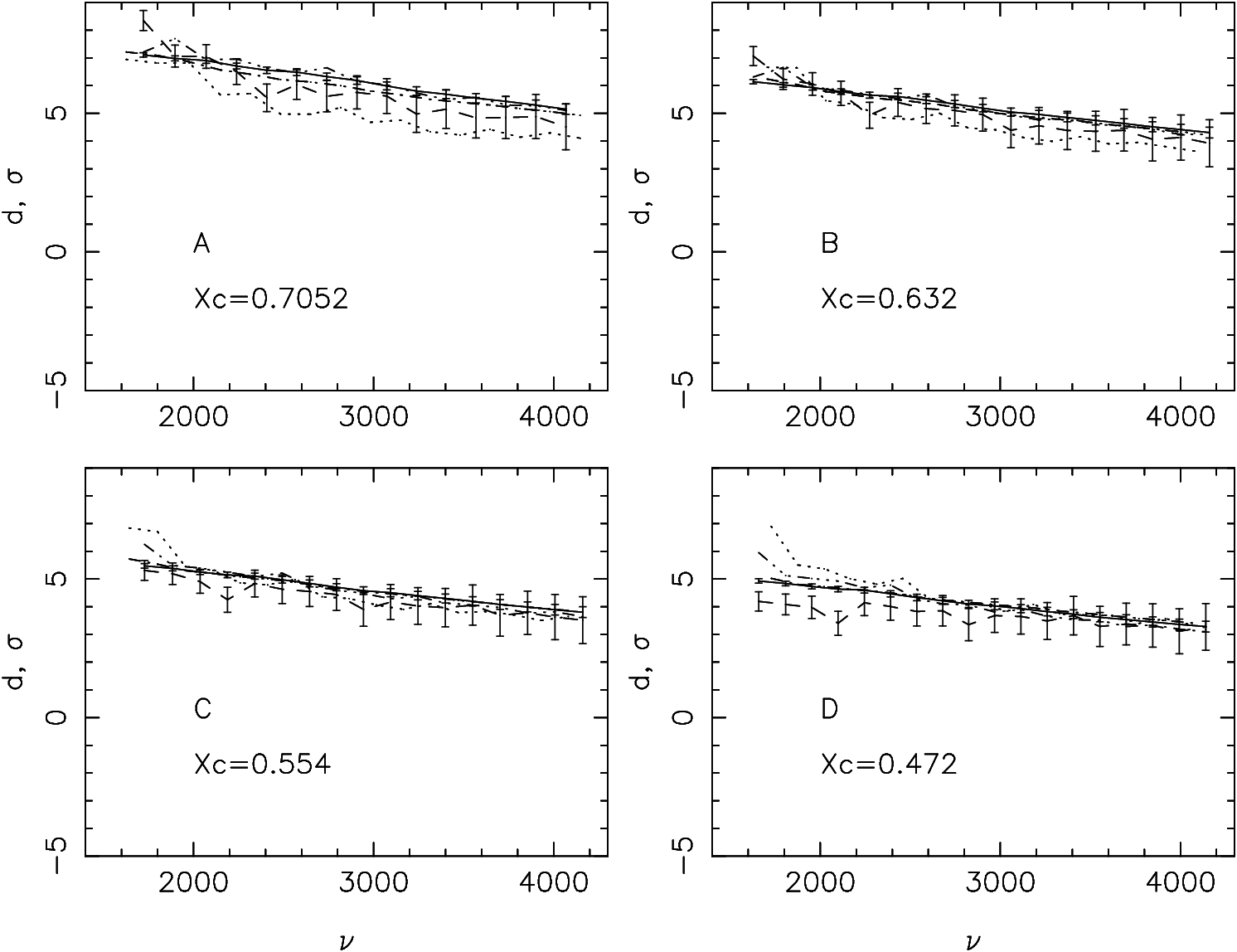}
     \includegraphics[angle=-90, width=9cm]{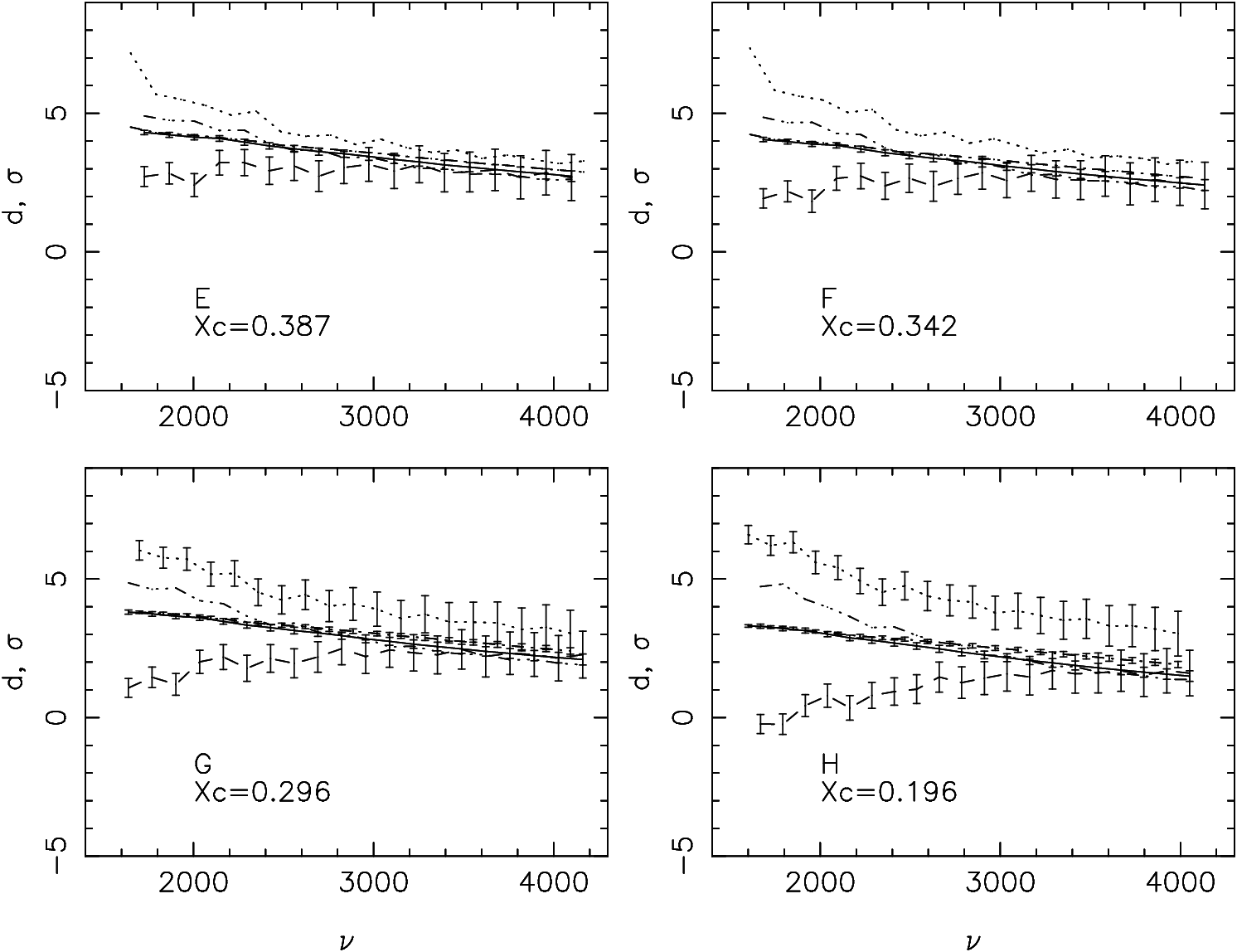}
     \includegraphics[angle=-90, width=9cm]{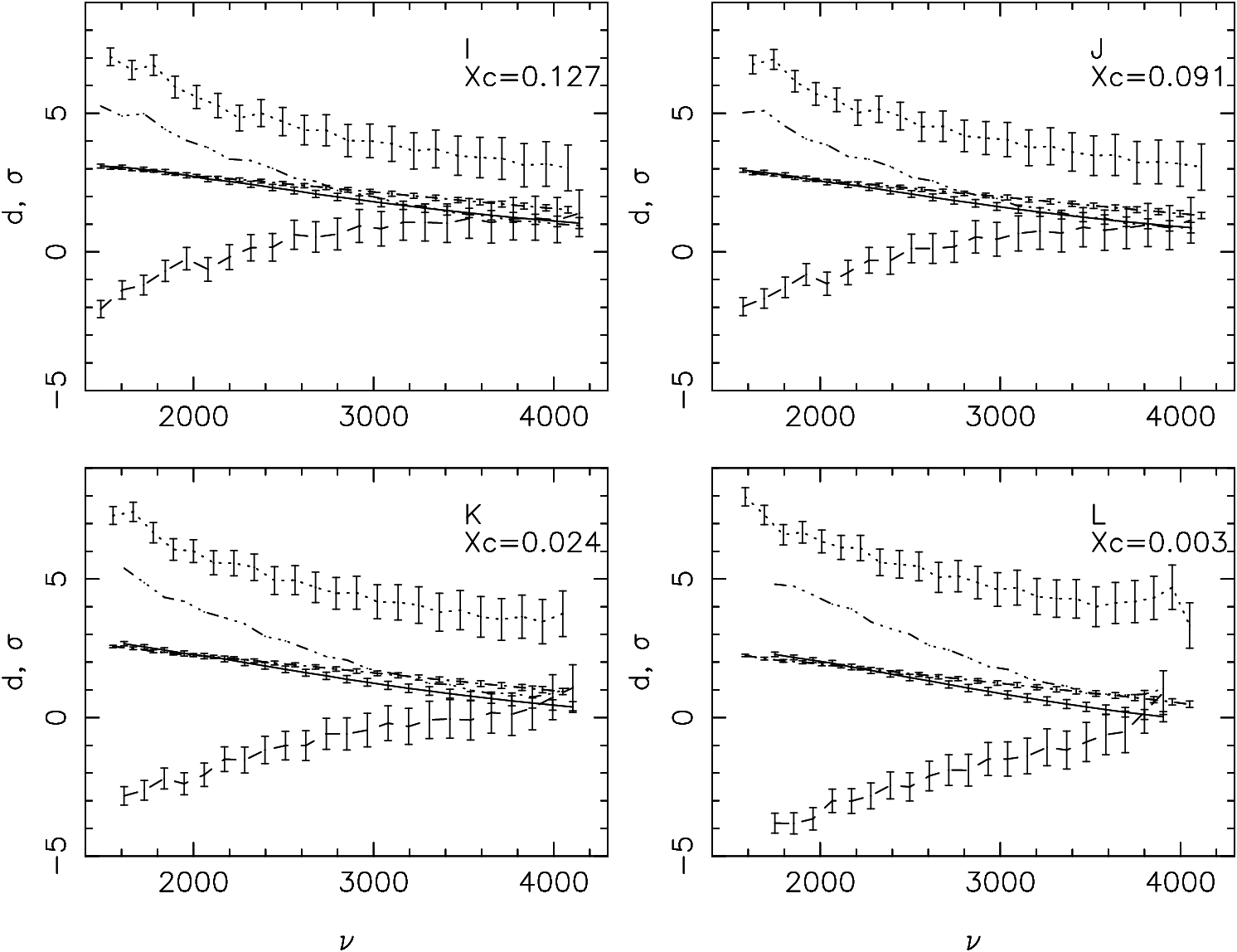}
     \centering
       \caption{Frequency separations $d_{02}(n)/3$: solid line,
       $d_{13}(n)/5$: dash-dotted line, $\sigma_{02}(n)$: dashed line,
       $\sigma_{13}(n)$: dotted line, and $d_{01}(n)$:
       triple-dot-dashed line as a function of frequencies $\nu_{n,l}$
       for model M1.0 at the different evolutionary stages. The
       errorbars represent 1$\sigma$ errors obtained assuming errors of
       1 part in $10^{4}$ in frequencies. For distinguishability, we plot
       only the errorbars of the $d_{02}/3$ and $\sigma_{02}$ between A and F.}
       \label{fig2}
   \end{figure*}

   In Fig. \ref{fig2}, we present the scaled small separations
   $d_{02}$/3 and $d_{13}$/5 and the differences $\sigma_{02}$
   and $\sigma_{13}$ computed from the numerically computed frequencies
   of model M1.0 as a function of frequency $\nu_{n, l}$ at the different
   evolutionary stage labeled by the central hydrogen mass fraction $X_{c}$.
   The errorbars indicate 1$\sigma$ errors obtained assuming errors
   of 1 part in $10^{4}$ in frequencies. In Table \ref{tab2}, we
   list the assumed errors of $\nu_{n,0}$ and the errors of $d_{02}(n)/3$ and
   $\sigma_{02}(n)$ of model M1.0 at the age of 4.5 Gyr. The errors
   in $\sigma_{02}(n)$ are larger than the errors in $d_{02}(n)/3$.
   At the early evolutionary stage, showed in Figs. \ref{fig2} A, B, and C,
   the differences $\sigma_{02}$ and $\sigma_{13}$ cannot be
   distinguished from the scaled small separations $d_{02}$/3 and $d_{13}$/5.
   From $X_{c} \sim$ 0.5 to $X_{c} \sim$ 0.003, the difference
   $\sigma_{l-1 l+1}(n)$ deviates more and more from the scaled small
   separation $d_{l l+2}(n)/(2l + 3)$; the $\sigma_{02}$ is less than
   the $d_{02}$/3 and $d_{13}$/5, but the $\sigma_{13}$ becomes larger
   than the $d_{02}$/3 and $d_{13}$/5. The less central the hydrogen, the more
   deviation. We also plot the quantity $d_{01}(n)$ in Fig. \ref{fig2}.
   The values of $d_{01}(n)$ are between the values of $\sigma_{13}(n)$ and
   of $d_{02}(n)/3$. With the decrease in $X_{c}$,
   the changes of the separation $\sigma_{13}$ in Fig. \ref{fig2} are small.
   But the separations $d_{l l+2}/(2l+3)$ and $\sigma_{02}$ in Fig. \ref{fig2}
   decrease with decrease in $X_{c}$. The difference $\sigma_{02}(n)$
   is somewhat more sensitive to the $X_{c}$ than the differences
   $d_{02}(n)/3$ and $d_{13}(n)/5$. In Fig. \ref{fig2}, the scaled
   small separation is smoother than the differences $\sigma_{02}(n)$,
   $\sigma_{13}(n)$, and $d_{01}(n)$. The scatter of the differences
   $\sigma_{02}(n)$ and $\sigma_{13}(n)$ may be related to $\sigma_{l-1 l+1}(n)$
   depending on the conditions not only in the stellar core but also
   in the envelope just as $d_{01}(n)$ (Oti Floranes et al. \cite{flo05}).

   \begin{table}
   \caption[]{Some of errors of model M1.0 at the age of 4.5 Gyr.}
   \label{tab2}
   \centering
   \begin{tabular}{c c c c c }
   \hline\hline
     n & $\nu_{n,0}$ &   $d_{02}(n)/3$ & $\sigma_{02}(n)$  \\
   \hline
11 & $ 1683.346 \pm 0.168 $ & $ 4.053 \pm 0.079 $ & $ 1.929 \pm0.349$ \\
12 & $ 1818.643 \pm 0.182 $ & $ 3.979 \pm 0.085 $ & $ 2.183 \pm0.376 $ \\
13 & $ 1954.024 \pm 0.195 $ & $ 3.895 \pm 0.092 $ & $ 1.828 \pm0.403 $ \\
14 & $ 2089.951 \pm 0.209 $ & $ 3.846 \pm 0.098 $ & $ 2.655 \pm0.431 $ \\
15 & $ 2225.364 \pm 0.223 $ & $ 3.712 \pm 0.105 $ & $ 2.737 \pm0.458 $ \\
16 & $ 2360.143 \pm 0.236 $ & $ 3.577 \pm 0.111 $ & $ 2.384 \pm0.485 $ \\
17 & $ 2494.034 \pm 0.249 $ & $ 3.482 \pm 0.117 $ & $ 2.657 \pm0.511 $ \\
18 & $ 2628.444 \pm 0.263 $ & $ 3.377 \pm 0.124 $ & $ 2.368 \pm0.538 $ \\
19 & $ 2763.917 \pm 0.276 $ & $ 3.300 \pm 0.130 $ & $ 2.648 \pm0.566 $ \\
20 & $ 2899.730 \pm 0.290 $ & $ 3.200 \pm 0.136 $ & $ 2.855 \pm0.593 $ \\
21 & $ 3036.027 \pm 0.304 $ & $ 3.088 \pm 0.143 $ & $ 2.575 \pm0.620 $ \\
22 & $ 3172.335 \pm 0.317 $ & $ 2.999 \pm 0.149 $ & $ 2.835 \pm0.647 $ \\
23 & $ 3308.821 \pm 0.331 $ & $ 2.896 \pm 0.156 $ & $ 2.617 \pm0.675 $ \\
24 & $ 3445.909 \pm 0.345 $ & $ 2.812 \pm 0.162 $ & $ 2.590 \pm0.702 $ \\
25 & $ 3583.206 \pm 0.358 $ & $ 2.730 \pm 0.169 $ & $ 2.741 \pm0.730 $ \\
26 & $ 3720.942 \pm 0.372 $ & $ 2.640 \pm 0.175 $ & $ 2.455 \pm0.757 $ \\
27 & $ 3858.897 \pm 0.386 $ & $ 2.568 \pm 0.182 $ & $ 2.606 \pm0.785 $ \\
28 & $ 3996.970 \pm 0.400 $ & $ 2.487 \pm 0.188 $ & $ 2.495 \pm
0.813 $\\

   \hline
   \end{tabular}
   \begin{list}{}{}
     \item Note.-- The errors of 1 part in $10^{4}$ in frequencies are assumed.
     The errors of $d_{02}/3$ and $\sigma_{02}$ are obtained using the
     error propagation formula and the assumed errors in frequencies.
   \end{list}
   \end{table}

\begin{figure*}
     \includegraphics[angle=-90, width=9cm]{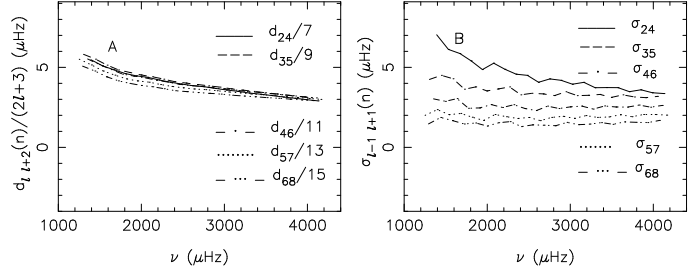}
     \centering
       \caption{(A) shows the small separations $d_{24}/7$, $d_{35}/9$,
       $d_{46}/11$, $d_{57}/13$, $d_{68}/15$ of model M1.0 at the stage
       of $X_{c}$ = 0.342. (B) shows the differences $\sigma_{24}$, $\sigma_{35}$,
       $\sigma_{46}$, $\sigma_{57}$,
       $\sigma_{68}$ of model M1.0 at the stage of $X_{c}$ = 0.342.}
       \label{fig3}
   \end{figure*}

   In Figs. \ref{fig3} A and B, we compare the scaled small
   separation $d_{l l+2}(n)/(2l+3)$ with the difference
   $\sigma_{l-1 l+1}(n)$ computed using the frequencies of degree as high
   as 8 at the evolutionary stage of $X_{c}$ = 0.342 of the model M1.0.
   The difference $\sigma_{l-1 l+1}(n)$ has the
   order of the scaled small separation $d_{l l+2}(n)/(2l+3)$.
   Moreover, on the one hand, for a given $l$ and $l \geq 4$,
   $\sigma_{l-1 l+1}(n)$ is almost invariant; on the other hand,
   $\sigma_{l-1 l+1}(n)$ is more dependent on the degree $l$ than
   $d_{l l+2}(n)/(2l+3)$. From Eqs. (\ref{eigd}) and
   (\ref{eigsigma}), one can find that the difference
   between $d_{l l+2}(n)/(2l+3)$ and $\sigma_{l-1 l+1}(n)$ is obvious.
   $d_{l l+2}(n)/(2l+3)$ and $\sigma_{l-1 l+1}(n)$ depend on the different
   phase shifts $\varphi_{l}$, which strongly depend on the degree $l$
   (Roxburgh \& Vorontsov \cite{rox94a}, \cite{rox00}, \cite{rox03}).

\begin{figure*}
     \includegraphics[angle=-90, width=9cm]{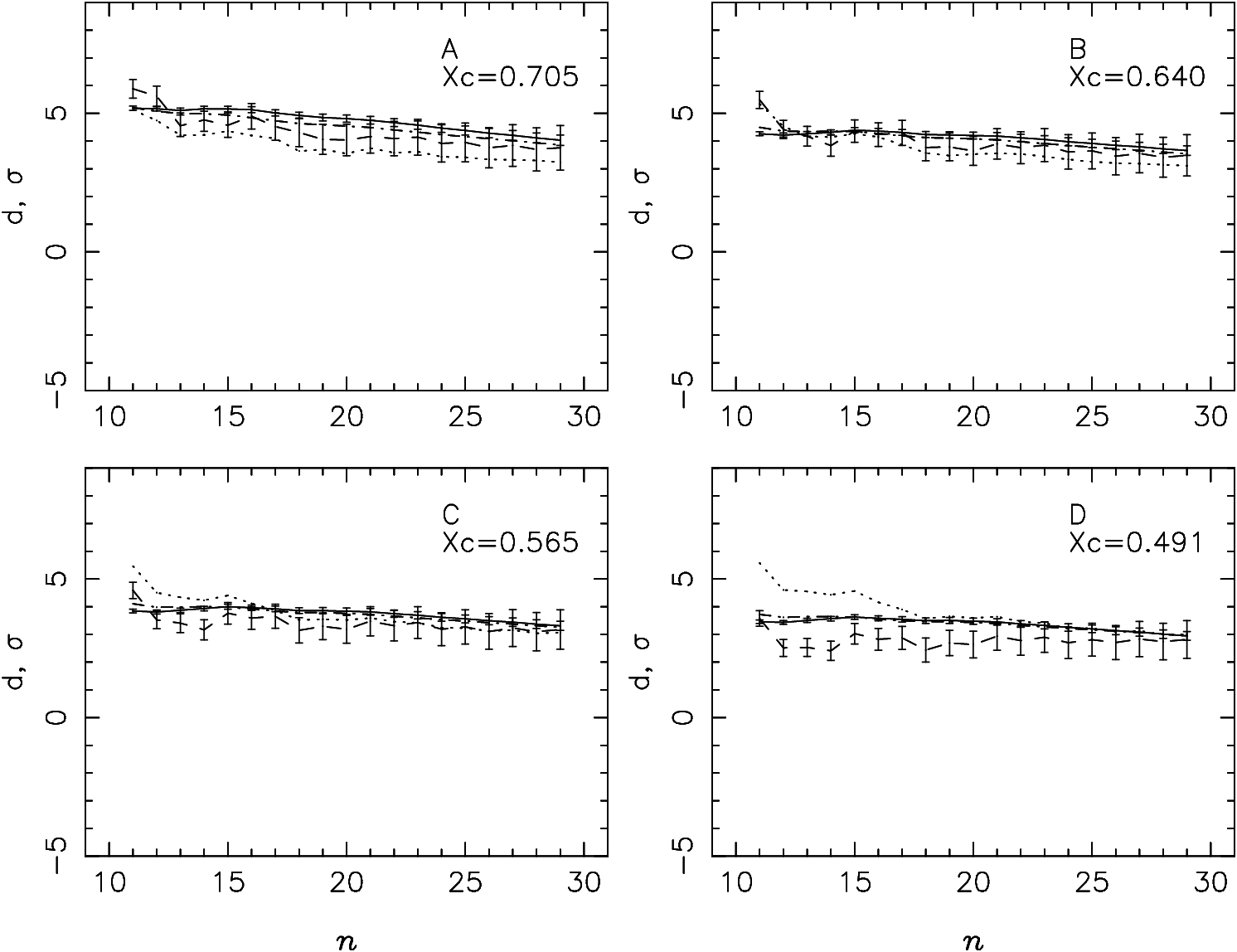}
     \includegraphics[angle=-90, width=9cm]{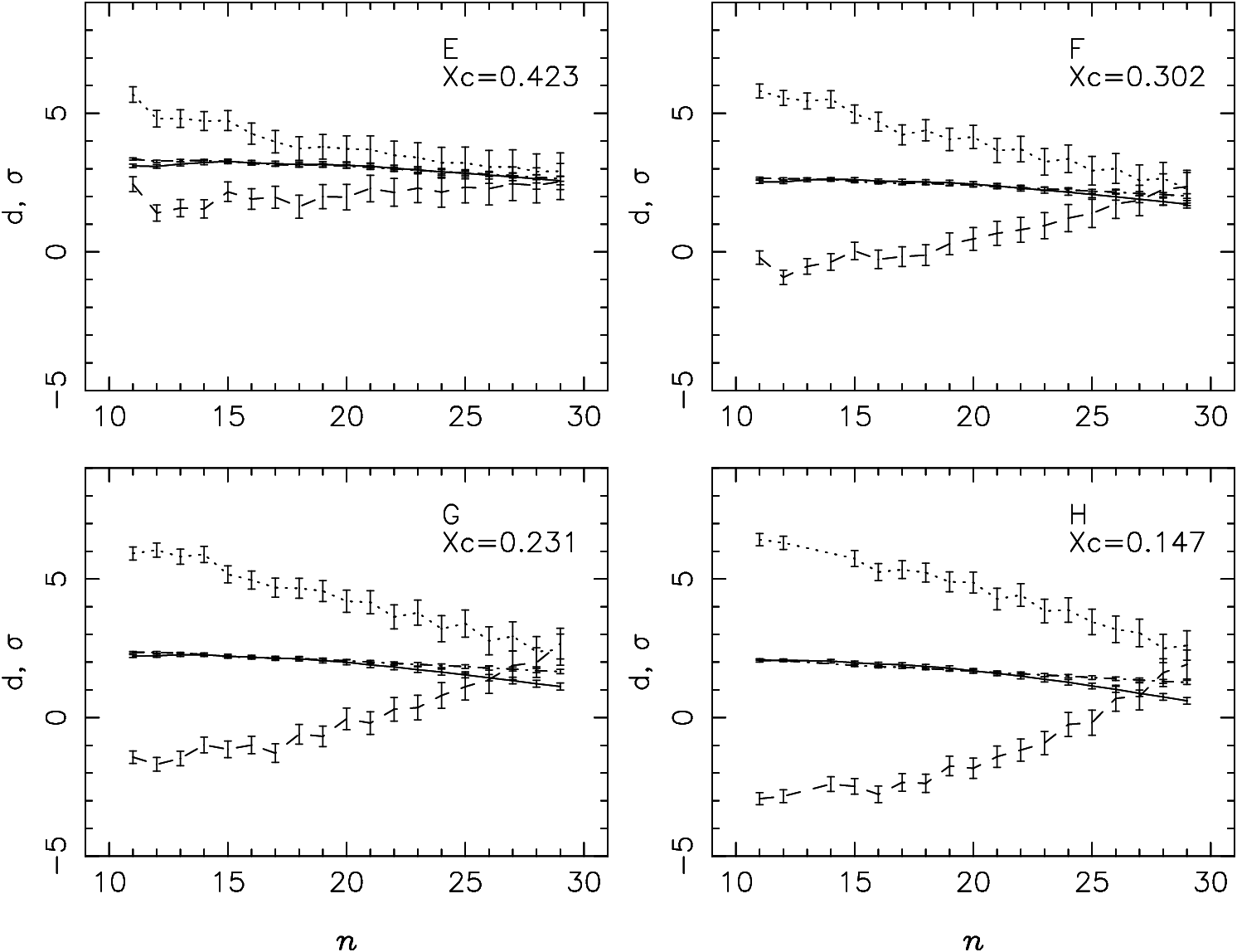}
     \centering
       \caption{Frequency separations $d_{02}(n)/3$: solid line,
       $d_{13}(n)/5$: dash-dotted line, $\sigma_{02}(n)$: dashed line,
       and $\sigma_{13}(n)$: dotted line as a function of order
       $n$ for model M1.2 at the different evolutionary stages.
       The errorbars represent 1$\sigma$ errors obtained
       assuming errors of 1 part in $10^{4}$ in frequencies.}
       \label{fig4}
   \end{figure*}

   Furthermore, in Fig. \ref{fig4}, we represent the separations
   $\sigma_{l-1 l+1}(n)$ and $d_{l l+2}(n)/(2l+3)$ of model M1.2 as a
   function of the order $n$. As results of the model M1.0,
   the separations $d_{ll+2}(n)/(2l+3)$ and $\sigma_{l-1 l+1}(n)$
   cannot be distinguished at the early evolutionary stage;
   with the decrease in $X_{c}$, the difference $\sigma_{l-1 l+1}(n)$
   deviates more and more from the scaled small separation too.
   For the models with $X_{c}<$ 0.423, the difference
   $\sigma_{l-1 l+1}(n)$ for $n < 25$ clearly deviates from the scaled
   small separation. The deviation between $\sigma_{l-1 l+1}(n)$ and
   $d_{l l+2}(n)/(2l+3)$ is related to order $n$.

\begin{figure*}
     \includegraphics[angle=-90, width=9cm]{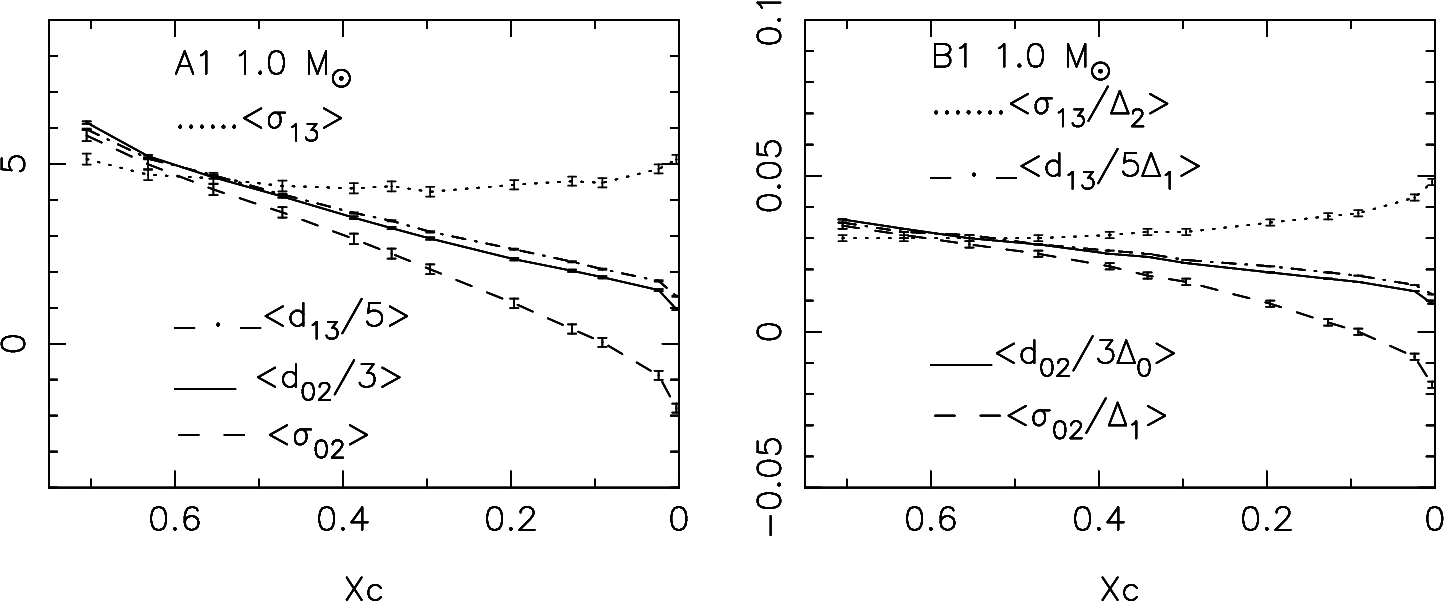}
     \includegraphics[angle=-90, width=9cm]{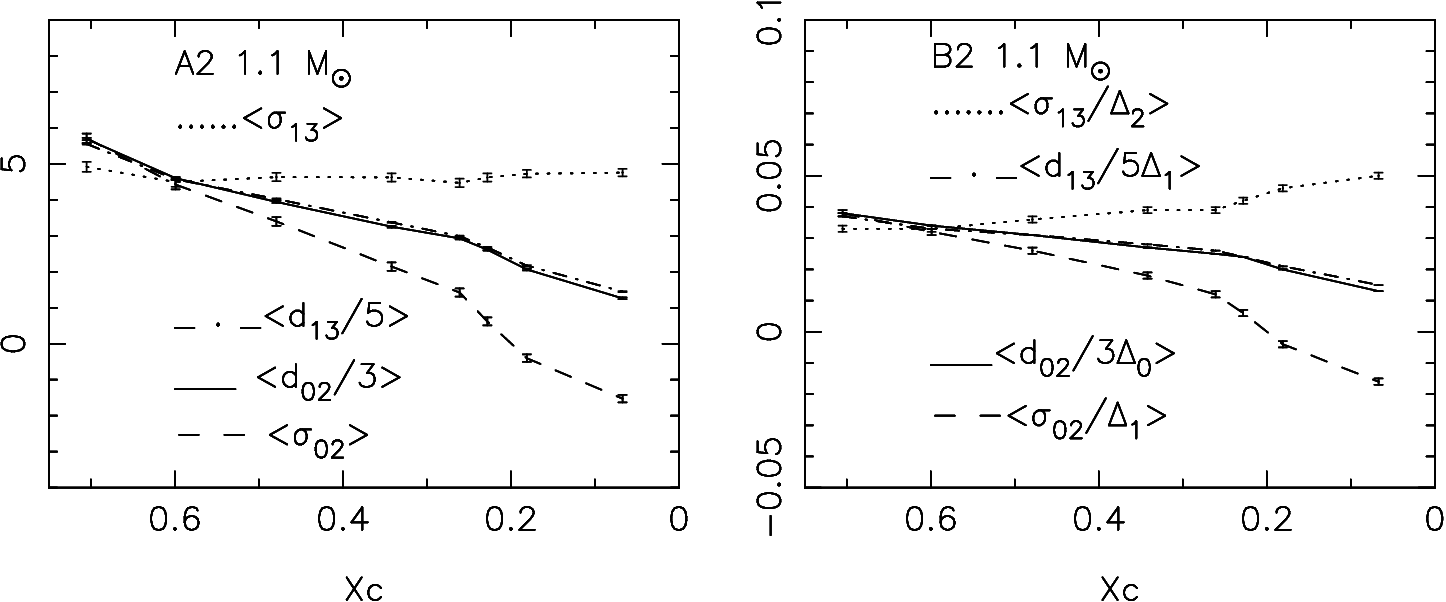}
     \includegraphics[angle=-90, width=9cm]{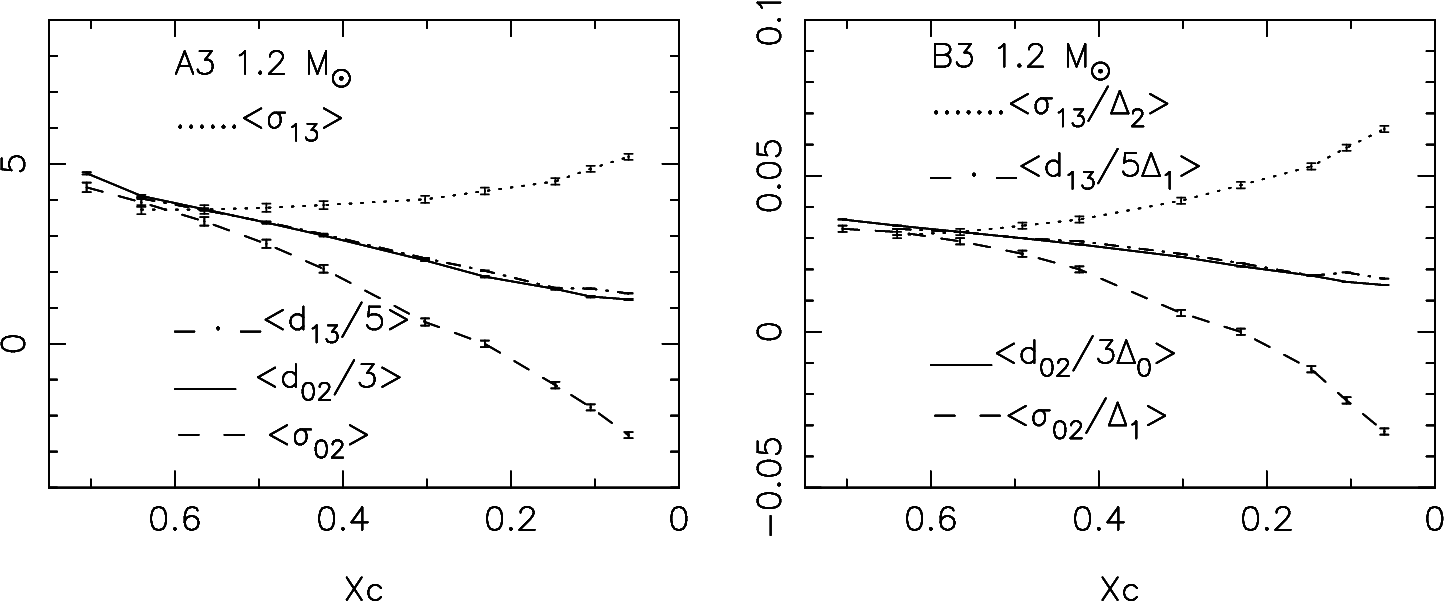}
     \centering
       \caption{\textbf{The average separations as a function of
       $X_{c}$. The errorbars represent
       1$\sigma$ errors obtained assuming errors of 1 part in $10^{4}$ in
       frequencies.}}
       \label{fig5}
   \end{figure*}

   In Figs. \ref{fig5} A1, A2, and A3, we depict the
   quantities $< d_{02}(n)/3>$, $<d_{13}(n)/5>$, $<\sigma_{02}(n)>$,
   and $<\sigma_{13}(n)>$ as a function of $X_{c}$, where $< >$
   implies averaging over a fixed range in frequencies or in orders.
   We choose the range of $1600 \leq \nu \leq 4100$ $\mu Hz$ for the model M1.0,
   $1200 \leq \nu \leq 3300$ $\mu Hz$ for the model M1.1, and 11 $\leq n\leq $ 29
   for the model M1.2. The choice of the range for averaging, as pointed
   out by Mazumdar et al. (\cite{maz06}), is governed by the availability
   of the data in the observation. It should be noted that the fixed domain
   in $n$ does not correspond to a fixed domain in frequency for models with
   different $X_{c}$. These ranges for averaging will be used in the rest
   of this paper. With the decrease in $X_{c}$, the average separations
   $<d_{02}/3>$, $<d_{13}/5>$, and $<\sigma_{02}>$ decrease; however,
   the average separation $<\sigma_{13}>$ increases slightly.
   Moreover, the average separation $<\sigma_{02}>$ is more sensitive to
   $X_{c}$ than the average separations $<d_{02}/3>$ and $<d_{13}/5>$.
   Thus the $<\sigma_{02}>$ and $<\sigma_{13}>$
   deviate more and more from the $<d_{02}/3>$ and $<d_{13}/5>$.

   It can be found from Eqs. (\ref{smalls}) and (\ref{sigs}) that the
   difference $\sigma_{l-1l+1}(n)$ and the scaled small separation
   $d_{l l+2}(n)/(2l+3)$ are dependent on the characteristic frequency
   $\nu_{0}$, which is affected by the outer layer of stars.
   The changes in $\nu_{0}$ must affect $\sigma_{l-1 l+1}(n)$ and
   $d_{l l+2}(n)/(2l+3)$. Noting that quantity $A$ is related to
   $\nu_{0}^{-1}$ in Eq. (\ref{eqa}) and the large separation is almost equal
   to $\nu_{0}$, we can use the large separation to eliminate the
   effect of $\nu_{0}$ on the difference $\sigma_{l-1 l+1}(n)$ and the
   scaled small separation. Thus, in Figs. \ref{fig5} B1, B2, and B3,
   we show the average ratio of scaled small separation to large
   separation, $<d_{l l+2}(n)/[(2l+3)\Delta_{l}(n)]>$, and the average ratio
   of the difference $\sigma_{l-1 l+1}(n)$ to large separation,
   $<\sigma_{l-1 l+1}(n)/\Delta_{l}(n)>$, as a function of the central
   hydrogen $X_{c}$. The value of $<\sigma_{l-1 l+1}(n)/\Delta_{l}(n)>$
   deviates from $<d_{l l+2}(n)/[(2l+3)\Delta_{l}(n)]>$ with the decrease
   in the central hydrogen abundance, as the behaviors between
   $<\sigma_{l-1 l+1}(n)>$ and $<d_{l l+2}(n)/(2l+3)>$. However,
   the $<\sigma_{13}/\Delta_{2}>$ obviously increases with the decrease
   in $X_{c}$ compared with the $<\sigma_{13}>$.

\section{Discussion and conclusions}
   Only a very limited number of modes $(l=0, 1, 2, 3)$ are likely to
   be observed in solar-like oscillations. Thus, our calculation
   mainly concentrated on these modes.

   Equations (\ref{smalls}) and (\ref{sigs}) hint that
   both the difference $\sigma_{l-1 l+1}(n)$ and the scaled small separation
   $d_{l l+1}(n)/(2l+3)$ are sensitive to the central conditions of stars,
   and they should have the same magnitude and characteristics.
   However, the expressions of the scaled small separation and of the
   $\sigma_{l-1 l+1}(n)$ are obtained from an approximative expression
   of frequency $\nu_{n, l}$, which is inaccurate (Roxburgh \& Vorontsov
   \cite{rox94a}; Audard \& Provost \cite{aud94}). Although
   Eqs. (\ref{smalls}) and (\ref{sigs}) can
   present some characteristics of the scaled small separation and
   the difference $\sigma_{l-1 l+1}(n)$, there must be some characteristics
   covered by Eqs. (\ref{smalls}) and (\ref{sigs}). From more accurate
   expressions (\ref{eigd}) and (\ref{eigsigma}), one can find that
   both $\sigma_{l-1 l+1}(n)$ and $d_{l l+1}(n)$ are mainly determined by
   the internal phase shifts $\varphi_{l}$, which only depends on the interior
   structure of the star (Roxburgh \& Vorontsov \cite{rox94a}, \cite{rox00},
   \cite{rox03}). The separations $\sigma_{l-1 l+1}(n)$ and $d_{l l+1}(n)/(2l+3)$
   depend on the different phase shifts, and $\sigma_{l-1 l+1}(n)$ must therefore
   be different from $d_{l l+1}(n)/(2l+3)$. Only under the approximation
   $\varphi_{l} \sim l(l+1)D_{\varphi}$, $\sigma_{l-1 l+1}(n)$ and
   $d_{l l+1}(n)/(2l+3)$ depend on the same quantities.

   The separation $\sigma_{l-1 l+1}(n)$ depends on the term
   $l(\nu_{n,l-1}-\nu_{n,l+1})/\nu_{n,l-1}$, neglected in Eq.
   (\ref{sigs}), which relies on order $n$ and degree $l$.
   Therefore, the separations $\sigma_{02}$ and $\sigma_{13}$ are more
   dependent on order $n$ than separation $d_{l l+2}(n)/(2l+3)$ and
   $\sigma_{02}$ is different from $\sigma_{13}$,
   which is shown in Figs. \ref{fig2} and \ref{fig4}.

   The separation $\sigma_{l-1 l+1}(n)$ is more uncertain than
   the scaled small separation $d_{l l+2}(n)/(2l+3)$. In Table
   \ref{tab2}, we show the errors of $d_{02}/3$ and $\sigma_{02}$
   obtained from assuming errors in frequencies. The errors of
   $\sigma_{02}(n)$ are 4 times larger than the errors of $d_{02}(n)/3$.
   At the early evolutionary stage, the difference $\sigma_{l-1 l+1}(n)$
   cannot be distinguished from the scaled small separations.
   With the decrease in central hydrogen, the difference $\sigma_{02}(n)$
   becomes smaller, but the $\sigma_{13}(n)$ is larger than the
   scaled small separation. At the late evolutionary stage, the difference
   $\sigma_{l-1 l+1}(n)$ can thus be distinguished from the scaled small
   separation except for the separations of the high-order frequencies.
   The separations $d_{02}/3$, $d_{13}/5$, and $\sigma_{02}$
   decrease with the decrease in $X_{c}$. These separations are good
   indicators of $X_{c}$.

   The separations $\sigma_{02}$ and $\sigma_{13}$
   are easily distinguished from the separations $d_{02}/3$
   and $d_{13}/5$ of model M1.2 in Fig. \ref{fig4} compared with those
   of model M1.0 in Fig. \ref{fig2}. There is a convective core in
   model M1.2. Furthermore, model M1.1 also has a convective core
   for $X_{c}\lesssim$ 0.261. In Fig. \ref{fig5} A2, the average separation
   $<\sigma_{02}>$ has an obvious change at $X_{c} \approx$ 0.261.
   Therefore the separation $\sigma_{02}$ may be sensitive to
   the convective core.

   The difference $\sigma_{l-1 l+1}(n)$ is similar to the scaled small
   separation $d_{l l+2}/(2l+3)$. They have the same asymptotic
   formula for the low-degree p-modes and are mainly determined by the
   conditions of stellar core. However, $\sigma_{l-1 l+1}(n)$ is
   somewhat different from $d_{l l+2}(n)/(2l+3)$, particularly the
   $\sigma_{02}(n)$ and $\sigma_{13}(n)$. With the decrease in
   the central hydrogen, $\sigma_{02}(n)$ and $\sigma_{13}(n)$
   deviate more and more from the scaled small separation. This
   characteristic provides us with a possibility for probing the central
   hydrogen abundance of stars.

-------------------------------------------

\begin{acknowledgements}
We thank the anonymous referee for his/her useful remarks and the
NSFC through projects 10473021 and 10433030.

\end{acknowledgements}


\begin{thebibliography}{}

\bibitem[1994]{ale94}
Alexander, D. R., Ferguson, J. W. 1994, ApJ, 437, 846
\bibitem[1994]{aud94} Audard, N., \& Provost, J. 1994, A\&A, 282, 73
\bibitem[2001]{bed01} Bedding, T. R., Butler, P. R., Kjeldsen, H.,
et al. 2001, ApJ, 549, L105
\bibitem[2004]{bed04} Bedding, T. R., Kjeldsen, H., Butler, P. R.,
et al. 2004, ApJ, 614, 380
\bibitem[2001]{bou01} Bouchy, F., \& Carrier, F. 2001, A\&A, 374, L5
\bibitem[2002]{bou02} Bouchy, F., \& Carrier, F. 2002, A\&A, 390,
205
\bibitem[2003]{car03} Carrier, F., Bourban, G. 2003, A\&A, 406, L23
\bibitem[2006]{cas06} Castro, M., \& Vauclair, S. 2006, A\&A, 456,
616
\bibitem[1984]{chris84} Christensen-Dalsgaard, J. 1984, in Mangeney
A., Praderie, F., eds, Space Research Prospects in Stellar Activity
and Variability, Paris Observatory Press, Paris, P11
\bibitem[1988]{chris88} Christensen-Dalsgaard, J. 1988, in Advances
in Helio and Asteroseismology, ed. J. Christensen-Dalsgaard, \& S.
Frandsen (Reidel), 295
\bibitem[1993]{chris93} Christensen-Dalsgaard, J. 1993, in Seismic
Investigation of the Sun and Stars, ed. T. M. Bron, A.S.P. Conf.
Ser, 42, 347
\bibitem[2004]{egg04} Eggenberger, P., Charbonnel, C, \& Talon, S.
et al. 2004, A\&A, 417, 235

\bibitem[1987]{gou87} Gough, D. O. 1987, Nature, 326, 257
\bibitem[1990]{gou90a} Gough, D. O., \& Novotny, E. 1990, Soph, 128, 143
\bibitem[1990]{gou90b} Gough, D. O. 1990, in Progress of
Seismology of the Sun and Stars, Proc. Oji International Seminar
Hakone (Japan: Springer Verlag), Lect. Notes Phys., 367, 283
\bibitem[2003]{gou03} Gough, D. O. 2003, AP\&SS, 284, 165
\bibitem[1992]{gue92} Guenther, D. B., Demarque, P., Kim, Y.-C.,
\& Pinsonneault, M. H. 1992, ApJ, 387, 372G
\bibitem[1996]{igl96} Iglesias, C., Rogers, F. J., 1996, ApJ, 464, 943
\bibitem[2003]{ker03} Kervella, P., Th\'{e}venin, F., S\'{e}gransan,
D. et al. 2003, A\&A, 404, 1087
\bibitem[1995]{kje95} Kjeldesn, H., \& Bedding, T, R. 1995, A\&A,
293, 87

\bibitem[1999]{mar99} Marti\'{c}, M., Schmitt, J., Lebrun, J. et al.
1999, A\&A, 351, 993
\bibitem[2005]{maz05} Mazumdar, A. 2005, A\&A, 441, 1079
\bibitem[2006]{maz06} Mazumdar, A., Basu, S., \& Collier, B. L. et
al. 2006, MNRAS, 372, 949
\bibitem[1998]{mon98} Monteiro, M. J. P. F. G., \& Thompson, M. J.
1998, in New Eyes to See Inside the Sun and Stars (Dordrecht:
Kluwer), ed. F. L. Deubner, J. Christensen-Dalsgaard, \& D. W.
Kurtz, Proc. IAU Symp., 185, 317
\bibitem[2005]{flo05}Oti Floranes, H., Christensen-Dalsgaard, J., \&
Thompson, M. J. 2005, MNRAS, 356, 671
\bibitem[2002]{pou02} Pourbaix, D., Nidever, D., McCarthy, C. et al.
2002, A\&A, 386, 280

\bibitem[2002]{rog02} Rogers, F. J., \& Nayfonov A, ApJ 2002, 576,1064
\bibitem[1993]{rox93} Roxburgh, I. W. 1993, in PRISMA, Report of
Phase A Study, ed. T. Approurchaux, et al., ESA 93, 31
\bibitem[1994a]{rox94a} Roxburgh, I. W., \& Vorontsov, S. V. 1994a,
MNRAS, 267, 297
\bibitem[1994b]{rox94b} Roxburgh, I. W., \& Vorontsov, S. V. 1994b,
MNRAS, 268, 143
\bibitem[2000]{rox00} Roxburgh, I. W., \& Vorontsov, S. V. 2000,
MNRAS, 317, 141
\bibitem[2001]{rox01} Roxburgh, I. W., \& Vorontsov, S. V. 2001, MNRAS,
322, 85
\bibitem[2003]{rox03} Roxburgh, I. W., \& Vorontsov, S. V. 2003,
A\&A, 411, 215
\bibitem[2005]{rox05} Roxburgh, I. W. 2005, A\&A, 434, 665

\bibitem[2001]{sch01} Schou, J., \& Buzasi, D. L. 2001, soho, 10, 391S
\bibitem[1980]{tas80} Tassoul, M. 1980, ApJS, 43, 469
\bibitem[2005]{the05} Th\'{e}ado, S., Vauclair, S., \& Castro, M.
et al. 2005, A\&A, 437, 553
\bibitem[2002]{the02} Th\'{e}venin, F., Provost, J., \& Morel, P. et al.
2002, A\&A, 392, L9
\bibitem[1994]{tho94} Thoul, A. A., Bahcall, J. N., Loeb, A. 1994, ApJ,
421, 828
\bibitem[1986]{ulr86} Ulrich, R. K. 1986, ApJL, 306, L37
\bibitem[1988]{ulr88} Ulrich, R. K. 1988, IAUS, 123, 2990
\bibitem[2004]{vau04} Vauclair, S., \& Th\'{e}ado, S. 2004, A\&A,
425, 179



\end{thebibliography}
\end{document}